\documentclass[aps,a4paper, preprint, superscriptaddress,preprintnumbers,floatfix,nofootinbib,amsmath,amssymb]{revtex4-1}

\usepackage{url}
\usepackage{hyperref}
\usepackage{color}
\usepackage{cancel}
\usepackage{cleveref}
\usepackage{subfig}
\usepackage{soul}
\usepackage[normalem]{ulem}

\usepackage{amstext,amssymb}
\usepackage{amsmath}
\usepackage{graphicx}
\usepackage{xspace}
\usepackage{color}
\usepackage{units}
\usepackage[T1]{fontenc}
\usepackage{amsmath,bm}
\usepackage{nicefrac}
\usepackage{slashed} % feynman slash notation via \slashed{p}
\usepackage{multirow}
\newcommand{\be}{\begin{equation}}
\newcommand{\ee}{\end{equation}}
\newcommand{\bea}{\begin{eqnarray}}
\newcommand{\eea}{\end{eqnarray}}

\newcommand{\emu}{\eta_{e\mu}}
\newcommand{\etau}{\eta_{e\tau}}
\newcommand{\mutau}{\eta_{\mu\tau}}
\newcommand{\mycomment}[1]{}

\begin{document}

\title{Impact of scalar NSI with off-diagonal parameters at DUNE and P2SO}

\author{Sambit Kumar Pusty}
\email{pustysambit@gmail.com}
\affiliation{School of Physics,  University of Hyderabad, Hyderabad - 500046,  India}

\author{Rudra Majhi}
\email{rudra.majhi95@gmail.com}
\affiliation{Nabarangpur College, Nabarangpur - 764059, Odisha, India}              
\author{Dinesh Kumar Singha}
\email{dinesh.sin.187@gmail.com}
\affiliation{School of Physics,  University of Hyderabad, Hyderabad - 500046,  India}

\author{Monojit Ghosh}
\email{mghosh@irb.hr}
\affiliation{Center of Excellence for Advanced Materials and Sensing Devices, Ru{\dj}er Bo\v{s}kovi\'c Institute, 10000 Zagreb, Croatia}             
\author{Rukmani Mohanta}
\email{rmsp@uohyd.ac.in}
\affiliation{School of Physics,  University of Hyderabad, Hyderabad - 500046,  India}

\begin{abstract}

In this paper, we studied the impact of the off-diagonal SNSI parameters in the future long-baseline neutrino oscillation experiments DUNE and P2SO. In our analysis, we found that the sensitivities of these experiments altered in a very non-trivial way due to the presence of these parameters. Depending on the values of these parameters, they can either completely mimic the standard scenario or can wash out their CP sensitivity. For large values of parameters $\eta_{e\mu}$ and $\eta_{e\tau}$, we obtained larger mass ordering and octant sensitivities as compared to the standard three flavour scenario. For the parameter $\eta_{\mu\tau}$, the mass ordering sensitivity and the precision of $\Delta m^2_{31}$ deteriorated compared to the standard scenario. Our results also showed that the sensitivities were significantly influenced by the phases of the off-diagonal parameters. 
\end{abstract}

\maketitle

\section{Introduction}

The origin of small and non-zero neutrino masses inferred from various neutrino oscillation experiments remains a mystery so far and thus provides an ideal platform to explore physics beyond the Standard Model (BSM).  Over the past few decades, there has been remarkable advancement in the precise determination of various neutrino oscillation parameters. Among these parameters, the mass-squared differences and the mixing angles, except $\theta_{23}$, are measured with a precision of a few percent. On the other hand, the CP-violating phase $\delta_{CP}$ remains uncertain so far. Additionally, the issues related to the sign of $\Delta m_{31}^2$, the so-called mass ordering problem, and the octant of $\theta_{23}$ are yet to be resolved~\cite{Esteban:2020cvm,Esteban:2024eli}. Despite these unknowns, the neutrino sector can be regarded as a means to probe new and unexplored areas of BSM physics. Neutrinos originating from various sources such as Earth's atmosphere, accelerators, and reactors, have been studied extensively in various oscillation experiments. In particular, long-baseline experiments have proven to be highly effective for the  precise determination of the oscillation parameters due to the presence of matter effects.  Consequently, these experiments provide an excellent opportunity to probe various BSM frameworks, which can induce sub-leading effects in the neutrino oscillation phenomena.  

Various sub-dominant repercussions of BSM Physics,  such as non-standard interactions (NSIs)~\cite{Antusch:2008tz,Gavela:2008ra,Ohlsson:2012kf,Biggio:2009nt,Farzan:2017xzy}, Lorentz invariance violation (LIV)~\cite{Greenberg:2002uu,Kostelecky:2003cr,Katori:2012pe,Diaz:2014yva}, long range force (LRF)~\cite{Joshipura:2003jh,Grifols:2003gy,Gonzalez-Garcia:2006vic}, neutrino decay~\cite{Bahcall:1972my,Lindner:2001fx,Beacom:2002cb,Bandyopadhyay:2002qg,Joshipura:2002fb}, decoherence \cite{Barenboim:2006xt,Lisi:2000zt,Fogli:2007tx,Guzzo:2014jbp}, etc., have been studied extensively in the context of long-baseline neutrino oscillation experiments. Within the realm of BSM physics, the NSI of neutrinos emerges as a notable and well-motivated phenomenon. Various experiments have already delved into  NSI signals,  associated with charge current (CC) \cite{Grossman:1995wx} as well as neutral current (NC)  \cite{Barger:1991ae} interactions. It should be emphasized that the CC and NC NSIs are mediated through vector fields. Corresponding effects of NC-NSI appeared as a potential term in the Hamiltonian for neutrino oscillation, whereas CC NSI modifies the neutrino flavour states in the time evolution equation. More generally, the CC-NSI affects the production and detection mechanisms of neutrinos at the source and detector,  while the NC-NSI affects the propagation of neutrinos between the source and detector. On the other hand, if the  NSIs are mediated by a scalar, the corresponding interactions contribute as a correction to the neutrino mass matrix rather than the matter potential. Consequently, the effect of scalar NSI becomes energy independent, while the vector one scales linearly with neutrino energy, which leads to significantly different phenomenological consequences in neutrino oscillation experiments. Recent studies elaborately discuss the NSI of neutrinos mediated by a light scalar particle ~\cite{Ge:2018uhz,Ge:2019tdi,Denton:2022pxt,Cordero:2022fwb,Gupta:2023wct,Medhi:2021wxj,Medhi:2022qmu,Medhi:2023ebi,Sarkar:2022ujy,Babu:2019iml,Singha:2023set,Dutta:2024hqq,ESSnuSB:2023lbg,Sarker:2023qzp,Sarker:2024ytu, Denton:2024upc, Bezboruah:2024yhk}. This newly emerged interaction is known as Scalar NSI (SNSI). Apart from providing additional contributions to the neutrino mass, SNSI can also worsen the determination of the unknowns in the neutrino sector. The effect of SNSI can be parameterised as a complex matrix with complex off-diagonal parameters and real diagonal parameters. Several articles have discussed the effects of diagonal and off-diagonal parameters at various long-baseline experiments, such as DUNE~\cite{Denton:2022pxt,Medhi:2021wxj,Medhi:2022qmu,Medhi:2023ebi,Sarkar:2022ujy,Dutta:2024hqq,Sarker:2023qzp,Sarker:2024ytu}, T2HK~\cite{Medhi:2022qmu,Sarker:2023qzp}, T2HKK~\cite{Medhi:2022qmu,Sarker:2023qzp}, ESSnuSB~\cite{Cordero:2022fwb,ESSnuSB:2023lbg} and P2SO~\cite{Singha:2023set}.

In one of our recent studies, we have explored the impact of the diagonal SNSI parameters in the context of DUNE and P2SO \cite{Singha:2023set}. In our work, we have shown that the atmospheric mass square difference $\Delta m^2_{31}$ plays a non-trivial role when one tries to put bounds on some of the diagonal SNSI parameters, and CP sensitivities of these experiments are lost for a particular value of the SNSI parameter $\eta_{ee}$. For the latter, we derived an analytical expression and showed this explicitly. In this paper, our aim is to extend that study by including the off-diagonal SNSI parameters. Unlike the diagonal parameters, the off-diagonal parameters are complex in nature and hence, there are additional phases associated with them. The main goal of this paper is to understand how these additional phases impact {(i) when one tries to put bounds on the SNSI parameters and (ii) when one tries to measure the sensitivities to the standard parameters. It will be interesting to see if the conclusions we obtained for the diagonal SNSI parameters remain the same for the off-diagonal NSI parameters. Further, we also expect to observe a significant change in the CP sensitivity, as the presence of the new phases can impact the sensitivity of the standard CP phase $\delta_{\rm CP}$ as discussed in the Appendix C of Ref.~\cite{Singha:2023set}. 

The paper is organized as follows. The next section will be on the theoretical background of SNSI parameters. Simulation details and experimental configuration will be discussed in the Section \ref{simulation}. Then in Section \ref{result}, we will present our results and findings on SNSI. Finally, we will conclude the paper with a summary in Section \ref{conclusion}.

\section{Theoretical Framework}
\label{theoretical-framework}

The Lagrangian corresponding to the simplest model that describes SNSI, can be expressed as \citep{Ge:2018uhz,Dutta:2024hqq},
\begin{eqnarray}
\mathcal{L} = \bar\nu_\alpha (i \gamma^\mu \partial_\mu - M_{ \alpha \beta}) \nu_\beta - (y_\nu)_{\alpha \beta} \bar{\nu}_{\alpha} \nu_\beta \phi - y_f \bar{f} f \phi - \frac{1}{2} (\partial_\mu \phi)^2 -\frac{m_\phi^2}{2} \phi^2
\label{lag}
\end{eqnarray}
where $y_f$ is the Yukawa coupling of the scalar mediator $\phi$ with fermion $f$, $y_\nu$ is the Yukawa coupling of the scalar mediator with the neutrinos $\nu$, and $m_\phi$ is the mass of the scalar mediator. Here $\alpha$ and $\beta$ are the flavour index of the leptons. Without delving into the details of  model building aspects, we presume that the  neutrinos interact with the fermions in a non-standard way through a scalar mediator. 

The corresponding effective Lagrangian can be written as,
\begin{eqnarray}
 \mathcal{L}_{\rm eff} = \frac{ y_{\alpha \beta} \hspace*{0.03 true cm} y_f}{m_\phi^2} (\overline{\nu}_\alpha\nu_\beta)(\bar f f ).
 \label{lag}
\end{eqnarray}
The Dirac equation in the presence of SNSI can be written as \cite{Ge:2018uhz}:
\begin{eqnarray}
\overline{\nu}_\alpha \left[i \partial_\mu \gamma^\mu + \left(M_{ \alpha \beta} + \frac{\sum_f N_f \hspace*{0.03 true cm} y_{\alpha \beta}\hspace*{0.03 true cm} y_f }{m_\phi^2}\right)\right] \nu_\beta = 0\;,
\label{dir}
\end{eqnarray}
where $M_{\beta \alpha}$ is the Dirac mass matrix of the neutrinos and $N_f$ is the number density of fermion $f$. Therefore, we note that the effect of SNSI appears as a correction term within the neutrino mass matrix. This correction can be characterized as
\begin{eqnarray}
\delta M =  \sqrt{|\Delta m^2_{31}|}
\begin{pmatrix}
\eta_{ee} & \eta_{e\mu} & \eta_{e\tau}\\
\eta_{\mu e} & \eta_{\mu \mu} & \eta_{\mu \tau} \\
\eta_{\tau e} & \eta_{\tau \mu} & \eta_{\tau\tau}
\label{eta}
\end{pmatrix}\;,
\end{eqnarray}
where we have considered $\eta_{\alpha\beta}$ as SNSI parameters. In order to make SNSI parameter $\eta_{\alpha\beta}$ dimensionless, we scale the magnitude of $\delta M$ relative to $\sqrt{|\Delta m^2_{31}|}$, where $\Delta m^2_{31} = m_3^2 - m_1^2$ is the atmospheric mass square difference. Comparing Eqs.~\ref{dir} and \ref{eta}, one can write
\begin{eqnarray}
    \eta_{\alpha \beta} = \frac{1}{m_\phi^2 \sqrt{|\Delta m^2_{31}|}} \sum_f N_f y_f y_{\alpha \beta}\;.
\label{eta_d}
\end{eqnarray}
 In order to have $\delta M$ to be Hermitian, we have considered the diagonal elements $\eta_{\beta \beta}$ as real and the off-diagonal elements $\eta_{\alpha \beta}$ with $\alpha \neq \beta$ as complex. As mentioned in the introduction, in this work we will focus only on the off-diagonal elements, which we parametrize as $\eta_{\alpha \beta} = |\eta_{\alpha \beta}| e^{i \phi_{\alpha \beta}}$. From Eq.~\ref{eta_d}, it is clear that the SNSI parameters depend on the matter density \footnote{Note that NSI mediated by a vector field (VNSI) also depends on the matter density. However, for VNSI, the matter term appears with energy $E$. To understand the separation of VNSI from SNSI at the Hamiltonian and probability level we refer to Appendix A of Ref.~\cite{Singha:2023set}}. As a result, care should be taken while comparing the values of $\eta_{\alpha\beta}$ in accordance with the matter density profile of different experiments. The matter densities for P2SO and DUNE experiments are nearly equal, and the results of $\eta_{\alpha \beta}$ can be comparable. 

At this point, it is important to comment on the present bounds on the Yukawa couplings and the mass of the scalar mediator and how our results can be correlated with these bounds. A detailed discussion on this topic is available in Refs.~\cite{Singha:2023set} and \cite{ESSnuSB:2023lbg}. In summary, the bounds obtained in our analysis are model independent. However, the effective couplings $\eta_{\alpha\beta}$ receive constraints from the experiments that are sensitive to neutrino-electron and/or neutrino-nucleon elastic scattering \citep{Dutta:2022fdt}. These bounds would also be expected to be model independent. As discussed in Ref.~\citep{Dutta:2024hqq}, this scenario is expected to be constrained from Borexino data on solar neutrinos and data from SN1987A. A discussion on constraints on SNSI parameters from solar and a combination of solar and reactor experiments is available in Ref. \cite{Denton:2024upc}. Finally, it has been shown in Ref.~\cite{Dutta:2022fdt} that the same values of the couplings satisfy a wide range of mediator mass. Therefore, there will not be a direct correlation between the bounds of the SNSI parameters obtained from scattering experiments and the neutrino oscillation experiments.

 The effective Hamiltonian of neutrino oscillation considering  scalar NSI can be written as
\begin{eqnarray}
  H = E_\nu + \frac{MM^\dagger}{2E_\nu} + V \;,
  \label{ham}
\end{eqnarray}
where $E_\nu$ is the energy of the neutrinos, $V = {\rm diag}(\sqrt{2}G_F N_e,0,0)$ is the standard matter potential with $G_F$ being the Fermi constant and $N_e$ represents the electron number density. In this case, the term $M$ becomes
\begin{eqnarray}
 M &=& U~{\rm diag}(m_1, m_2, m_3)~U^\dagger + \delta M \nonumber\\
   &=& U~{\rm diag}\left(m_1, \sqrt{m_1^2 + \Delta m^2_{21}}, \sqrt{m_1^2 + \Delta m^2_{31}}\right)~U^\dagger + \delta M\;,\label{h} 
 \end{eqnarray}
where we have assumed normal ordering of the neutrino masses i.e., $m_3 \gg m_2 > m_1$. Here $\Delta m^2_{21} = m_2^2 - m_1^2$ is the solar mass squared difference and $U$ is the standard PMNS matrix.  Neutrino oscillation probabilities in presence of SNSI can be calculated by diagonalizing Eq.~\ref{ham}. The neutrino oscillation probabilities will depend on the SNSI parameter as well as on the lightest neutrino mass $m_1$.

\section{Simulation Details}
\label{simulation}

In this section, we will discuss the details of the experimental configuration as well as the simulation methods used in this analysis. We have considered two long-baseline experiments DUNE and P2SO. The detailed configurations are outlined below.

\subsection*{P2SO}

The Protvino to Super-ORCA (P2SO) is an upcoming long-baseline experiment. In this setup, neutrinos will be generated at a U-70 synchrotron located in Protvino, Russia, and will travel to a detector positioned 2595 km away in the Mediterranean Sea, 40 km offshore Toulon, France. Detailed configuration information for the P2SO experiment can be found in Refs.~\cite{Akindinov:2019flp, Singha:2022btw, Majhi:2022fed}.

The accelerator is designed to produce a 450 kW beam, equating to $4 \times 10^{20}$ protons on target (PoT) annually for the P2SO configuration. Notably, the Super-ORCA detector employs a density that is ten times greater than the ORCA detector. The energy window for the P2SO experiment spans from 0.2 GeV to 10 GeV. The total run period considered is six years, divided into three years each for neutrino and antineutrino modes.
\subsection*{DUNE}

DUNE is an upcoming long-baseline neutrino oscillation experiment situated at Fermilab in the USA. We have used the official GLoBES files corresponding to the technical design report (TDR) \cite{DUNE:2021cuw} for simulating the DUNE experiment. The DUNE experiment operates over a broad range of neutrino energies, corresponding to a high beam power of 1.2 MW. The near detector is located at Fermilab, and the far detector is 40 kt liquid argon time projection chamber (LArTPC) situated in South Dakota.

In our analysis, we consider the total run-time for DUNE to be 13 years, comprising 6.5 years in neutrino mode and 6.5 years in antineutrino mode. This duration corresponds to an accumulation of $1.1 \times 10^{21}$ PoT per year. This configuration of DUNE corresponds to ten years of data gathering by the experiment under the nominal staging assumptions outlined in Ref. \cite{DUNE:2020jqi}.

For simulation, we are using GLoBES software package \cite{Huber:2004ka, Huber:2007ji}. In order to implement SNSI, we have modified the probability engine of GLoBES.
We estimate the sensitivity in terms of $\chi^2$ analysis. We use the Poisson log-likelihood and assume that it is $\chi^2$-distributed:
\begin{equation}
 \chi^2_{{\rm stat}} = 2 \sum_{i=1}^n \bigg[ N^{{\rm test}}_i - N^{{\rm true}}_i - N^{{\rm true}}_i \log\bigg(\frac{N^{{\rm test}}_i}{N^{{\rm true}}_i}\bigg) \bigg]\,,
\end{equation}
where $N^{{\rm test}}$ and $N^{{\rm true}}$ are the number of events in the test and true spectra respectively, and $n$ is the number of energy bins. The systematic error is incorporated by the method of pull~\citep{Fogli:2002pt,Huber:2002mx}. The systematic uncertainties taken into consideration in our analysis for the P2SO and DUNE experiments are displayed in Table \ref{table_sys}. Both background and signal normalization and shape errors have been taken into account. There are no shape errors included in the analysis for the DUNE experiment. The values of the oscillation parameters are taken from NuFit 6.0 \cite{Esteban:2024eli} and are listed in Table~\ref{tab:osc-parameters}. In our calculation, we have minimized over the parameters $\theta_{23}$, $\Delta m^2_{31}$, $\delta_{\rm CP}$ and the phases of the complex SNSI parameters appropriately. Unless specified, all other parameters have not been minimized. We present our results for the normal ordering of the neutrino masses and taken $m_1 = 10^{-5}$ eV. Throughout our analysis, we will consider only one SNSI parameter at a time.  

\begin{table} 
\centering
\begin{tabular}{|c|c|c|} \hline
Systematics     & P2SO          & DUNE  \\ \hline
Sg-norm $\nu_{e}$   & 5$\%$   & 2$\%$      \\ 
Sg-norm $\nu_{\mu}$    & 5$\%$            & 5$\%$ \\ 
Bg-norm    & 12$\%$     & 5$\%$ to 20$\%$\\ 
Sg-shape      & 11$\%$     & -\\ 
Bg-shape     & 4\% to 11$\%$       & - \\ 
\hline
\end{tabular}
\caption{Systematic errors and their values considered in our analysis. We have mentioned normalization error as ``norm" ,  signal as ``Sg" and background as ``Bg".}
\label{table_sys}
\end{table}

\begin{table}[htbp!]
    \centering
    \begin{tabular}{|c|c|c|} \hline 
         Parameters&  Best fit value $\pm 1\sigma$& $3\sigma$\\ \hline 
         $\sin^2\theta_{12}$&  $0.308^{+0.012}_{-0.011}$& $0.275 \to 0.345$\\ \hline 
         $\sin^2\theta_{13}$&  $0.02215^{+0.00056}_{-0.00058}$& $0.02030 \to 0.02388$\\ \hline 
         $\sin^2\theta_{23}$&  $0.47^{+0.017}_{-0.013}$& $0.435 \to 0.585$\\ \hline 
         $\delta_{CP} / ^\circ$&  $212^{+0.26}_{-0.41}$& $124 \to 364$\\ \hline 
         $\Delta m_{21}^2/ 10^{-5}~{\rm eV}^2$&  $7.49^{+0.19}_{-0.19}$& $6.92 \to 8.05$\\ \hline 
         $\Delta m_{31}^2/ 10^{-3}~{\rm eV}^2$&  $+2.513^{+0.021}_{-0.019}$& $+2.451 \to +2.578$\\ \hline
    \end{tabular}
    \caption{Oscillation parameters values with their $1\sigma$ and $3\sigma$ regions as mentioned in NuFIT 6.0. \cite{Esteban:2024eli}} 
    \label{tab:osc-parameters}
\end{table}

\section{Results}
\label{result}

In this section, we will show the impact of off-diagonal SNSI parameters on the sensitivities of P2SO and DUNE. First, we will estimate the capability of these experiments to put bounds on the SNSI parameters by taking the standard scenario in true and the SNSI in test and then we will study how the sensitivities corresponding to the standard oscillation parameters are altered in the presence of off-diagonal SNSI parameters by taking SNSI in both true and test. 

\subsection{Sensitivity limits on SNSI parameters}

In Ref.~\cite{Singha:2023set}, we have seen that the parameter $\Delta m^2_{31}$ plays a non-trivial role in constraining the diagonal SNSI parameters. To understand the situation for the off-diagonal SNSI parameters, in Fig.~\ref{fig:eta-ldm}, we have plotted the 90\% C.L. contours in $|\eta_{\alpha \beta}|$ vs $\Delta m^2_{31}$ plane. The left/middle/right panel is for $\eta_{e\mu}/\eta_{e\tau}/\eta_{\mu\tau}$. In each panel, the red contour is for DUNE and the green contour is for P2SO and the dark purple contour is for the combination of DUNE and P2SO. The y-axis range in the figure shows the allowed region of $\Delta m^2_{31}$ in its current $3 \sigma$ C.L. 
\begin{figure}
    \centering

     \includegraphics[width=168mm, height=62mm]{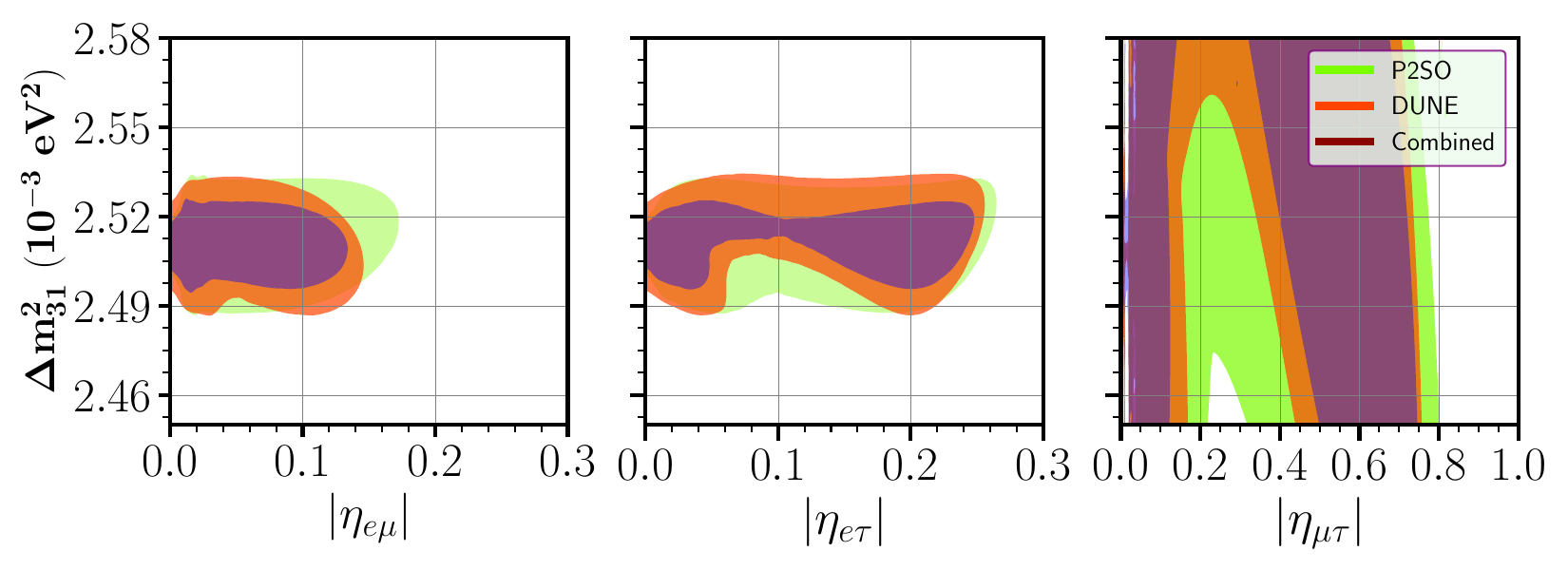}
     \caption{Allowed parameter space between $|\eta_{\alpha\beta}|$ and $\Delta m^2_{31}$ at 90\% C.L. for DUNE and P2SO experiment.  }
    \label{fig:eta-ldm}
\end{figure}
From this figure, we observed that for  $\eta_{e\mu}$ and $\eta_{e\tau}$, SNSI can be fitted with the standard scenario with a value of $\Delta m^2_{31}$ lying within its $3 \sigma$ region. This is evident from the close nature of the contours. This is in sharp contrast with the diagonal parameters $\eta_{\mu\mu}$ and $\eta_{\tau\tau}$, where the standard case can also be fitted with SNSI with a value of $\Delta m^2_{31}$ outside its $3 \sigma$ allowed values as shown in Ref.~\cite{Singha:2023set}. This means for the off-diagonal parameters $\eta_{e\mu}$ and $\eta_{e\tau}$, it will be sufficient to minimize $\Delta m^2_{31}$ within its current $3 \sigma$ range to obtain the correct upper bounds on these parameters. The scenario is slightly interesting for $\eta_{\mu\tau}$. For this parameter, we obtain an open-ended contour, implying that SNSI can mimic the standard scenario with a value of $\Delta m^2_{31}$ outside its current $3 \sigma$ values. However, we checked that the change in the sensitivities while estimating the upper bounds in both the cases i.e., when $\Delta m^2_{31}$ is minimized within its $3 \sigma$ values vs the case when $\Delta m^2_{31}$ minimized with a flat prior, are very similar. For this reason, we have chosen to vary this parameter within its $3 \sigma$ values throughout our analysis. From this figure we also see that the allowed parameter space for DUNE is more constrained compared to P2SO experiment. The combined analysis of DUNE and P2SO further constrains the allowed region, implying the fact that combined data from different experiments can enhance the sensitivities of the individual experiments. Note that one can further constrain the allowed regions between $|\eta_{\alpha\beta}|$ and $\Delta m^2_{31}$ by combining atmospheric neutrino oscillation experiments with the long-baseline neutrino experiment as discussed in Ref.\cite{Gratieri:2014jsa}. 

\begin{figure}[htbp!]
    \centering

     \includegraphics[width=168mm, height=62mm]{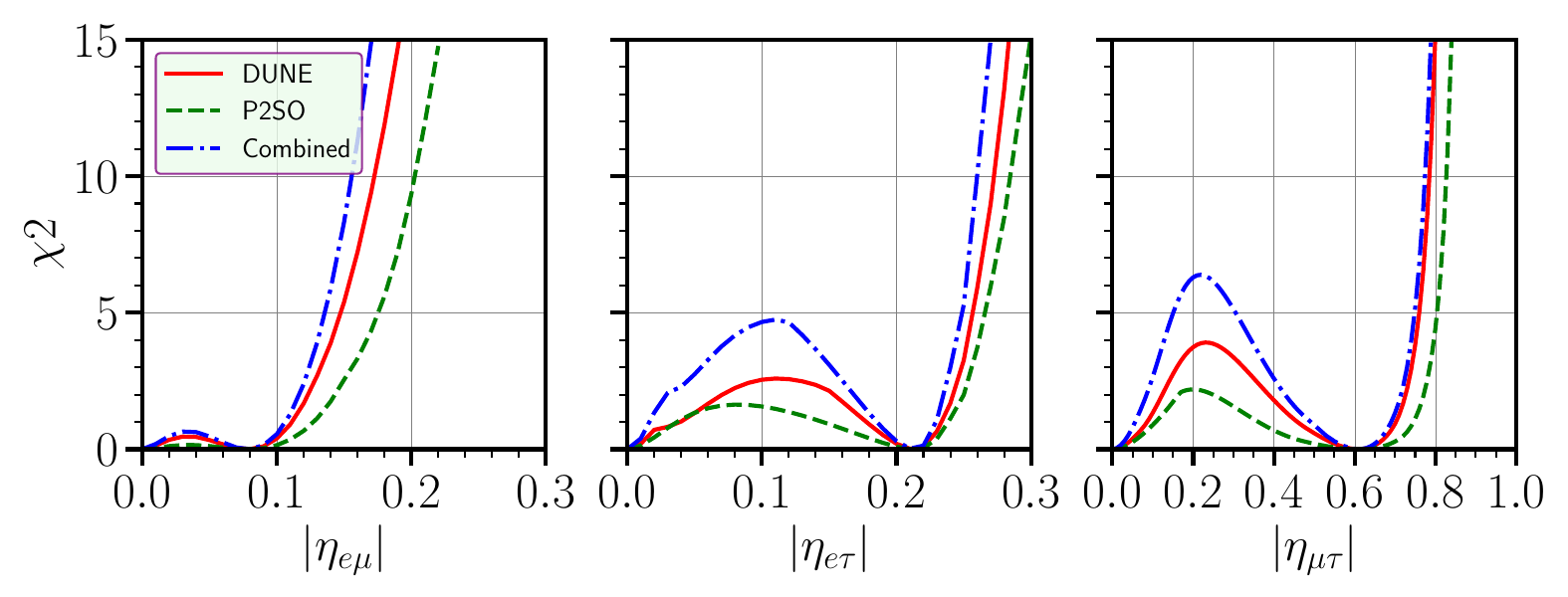}
    \includegraphics[width=70mm, height=60mm]{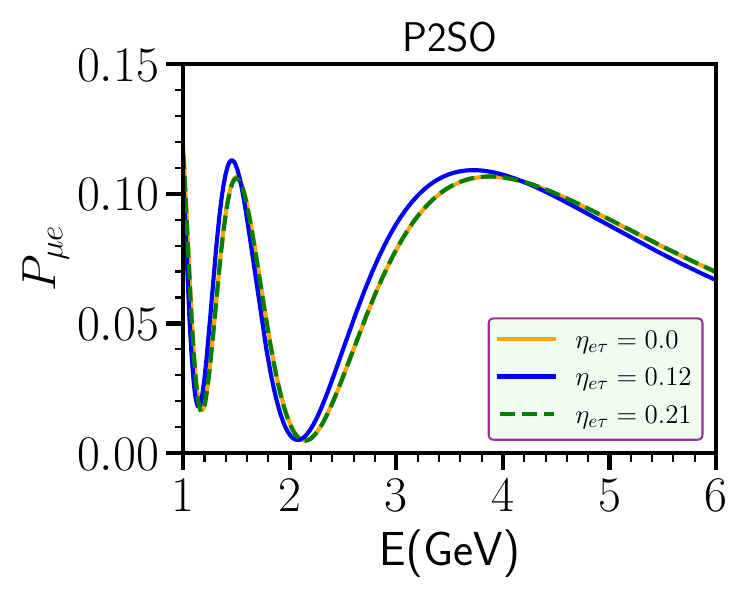} 
    \includegraphics[width=70mm, height=60mm]{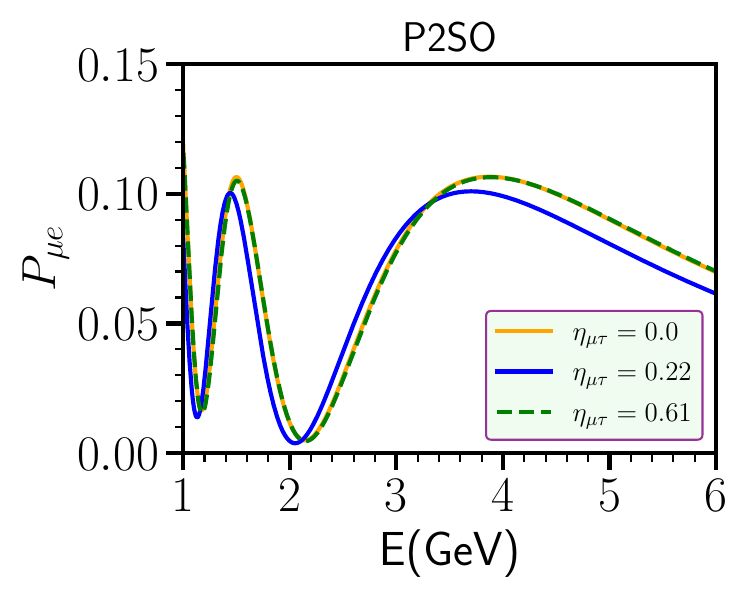} 
 \caption{Top row: Sensitivity limit on the SNSI parameters for P2SO and DUNE experiments. Bottom row: Appearance probability as a function of energy in P2SO for three different cases.}
    \label{fig:1D-bounds}
\end{figure}

In the top row of Fig.~\ref{fig:1D-bounds}, we have shown the sensitivity limits on the off-diagonal SNSI parameters for both P2SO and DUNE experiments. In each panel, the red curve represents the sensitivity for DUNE experiment while the green curve is for P2SO. We have shown a combined sensitivity of these two experiments, which is represented by the blue curve. From the panels, we see that the bound on $|\eta_{e\mu}|$ is stronger than $|\eta_{e\tau}|$, and the bound on $|\eta_{\mu\tau}|$ is the weakest among all. The sensitivities of DUNE and P2SO are very similar and the combined analysis of DUNE and P2SO can enhance the sensitivity on SNSI parameters. It is very interesting to note that all these three off-diagonal parameters mimic the standard scenario even when they assume non-zero values. For $\eta_{e\mu}/\eta_{e\tau}/\eta_{\mu\tau}$, this degeneracy appears at $0.08/0.21/0.61$, respectively. This degeneracy can be clearly understood from the lower panels of Fig. \ref{fig:1D-bounds}, where we have shown the appearance channel probability as a function of energy $E$ for P2SO. Left panel is for $\eta_{e\tau}$ and the right panel is for $\eta_{\mu\tau}$. The orange curve corresponds to the standard scenario and the green dashed curve corresponds to the value of the SNSI parameters for which the degeneracy occurs. The degeneracy is reflected by the fact that these two curves are completely overlapping. The blue curve, where there is no degeneracy, is separated from the other two curves. Similar results can also be obtained for the DUNE experiment.

\begin{figure}
   % \centering
     \includegraphics[width=168mm, height=62mm]{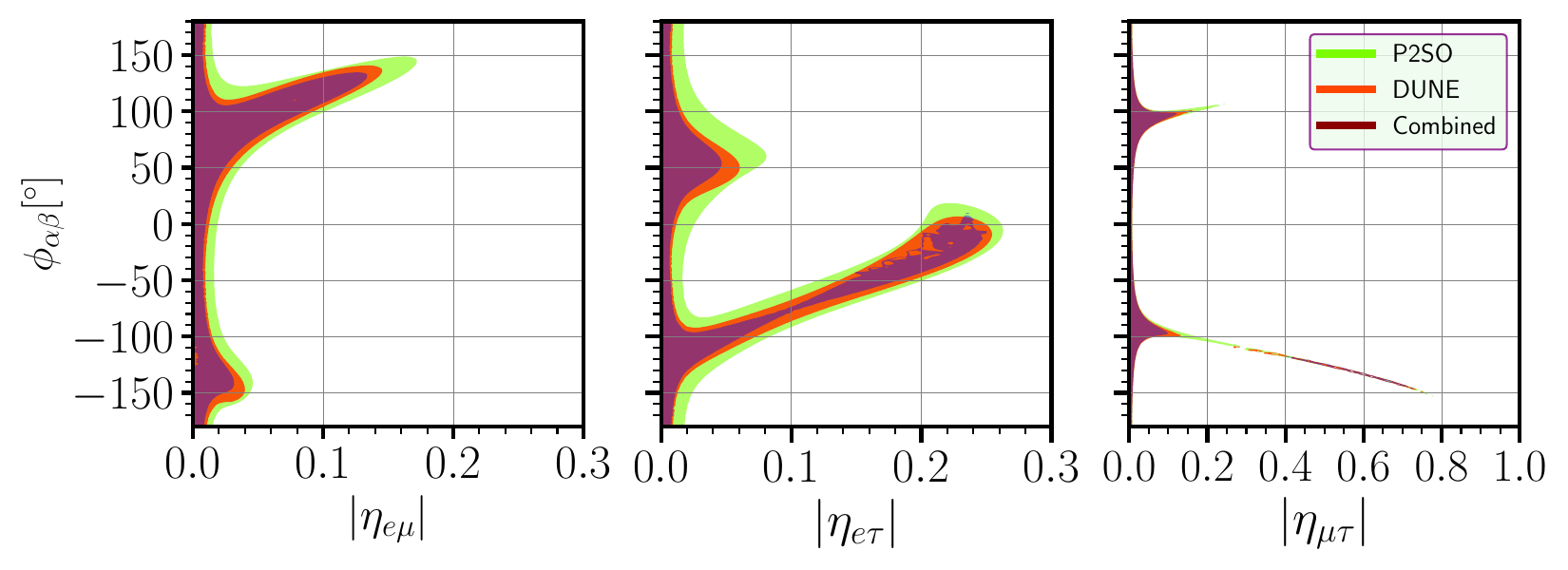}
    \caption{Allowed parameter space between SNSI parameter $|\eta_{\alpha\beta}|-\phi_{\alpha\beta}$ plane for P2SO and DUNE experiments at 90\% C.L. }
    \label{fig:eta-phi}
\end{figure}

\begin{table}[htbp!]
\begin{tabular}{|c|c|c|c|c|c|c|}
\hline 
\multirow{3}{*}{SNSI parameter} & 
\multicolumn{2}{c|}{DUNE} & 
\multicolumn{2}{c|}{P2SO} & \multicolumn{2}{c|}{DUNE+P2SO} \\ \cline{2-7} 
& Value & Phase ($\phi$) & Value & Phase ($\phi$) & Value & Phase ($\phi$) \\ 
\hline $\eta_{e\mu}$ & 0.15 & $136^{\circ}$ & 0.17 & $145^{\circ}$ & 0.13 & $132^{\circ}$ \\ 
\hline $\eta_{e\tau}$ & 0.25 & $-9^{\circ}$ & 0.26 & $-6^{\circ}$ & 0.247 & $-9.5^{\circ}$ \\ 
\hline $\eta_{\mu\tau}$ & 0.75 & $-149^{\circ}$ & 0.78 & $-152^{\circ}$ & 0.73 & $-146.5^{\circ}$ \\ 
\hline 
\end{tabular}
  \caption{Sensitivity limits on the SNSI parameters at 90\% C.L. from DUNE and P2SO experiments.}
  \label{tab:2}
\end{table}

To understand the role of the phases of the off-diagonal SNSI parameters, we have shown the upper bounds in the $|\eta_{\alpha\beta}|$ - $\phi_{\alpha\beta}$ plane at 90\% C.L. in Fig.~\ref{fig:eta-phi}. In each panel, the red (green) colored contour is the allowed parameter space for DUNE (P2SO) experiment and the dark purple colored contour is for the combination of DUNE and P2SO. In these panels, the degeneracies that we see} in Figs.~\ref{fig:eta-ldm} and \ref{fig:1D-bounds} are also visible. From these panels, we understand that the sensitivity of the off-diagonal parameters depends on the values of the associated phases. For example, in the case of $\eta_{e\mu}$, the sensitivity is weak around $\phi_{e\mu} = 136^\circ$, where we obtain our upper bound. However, if $\phi_{e\mu}$ happens to be around $0$, then the upper bound of $\eta_{e\mu}$ will be much stronger. In Table~\ref{tab:2}, we have listed the upper bounds of the off-diagonal SNSI parameters and their corresponding values of the phases, as obtained from DUNE and P2SO.

\subsection{Mass ordering sensitivity}
 %%%%%%%%%%%%%%%%%%%%%%%%%%%%%%%%%%%%%%%%%%%%%%%%
\begin{figure}[htbp!]
    \centering

  \includegraphics[width=168mm, height=62mm]{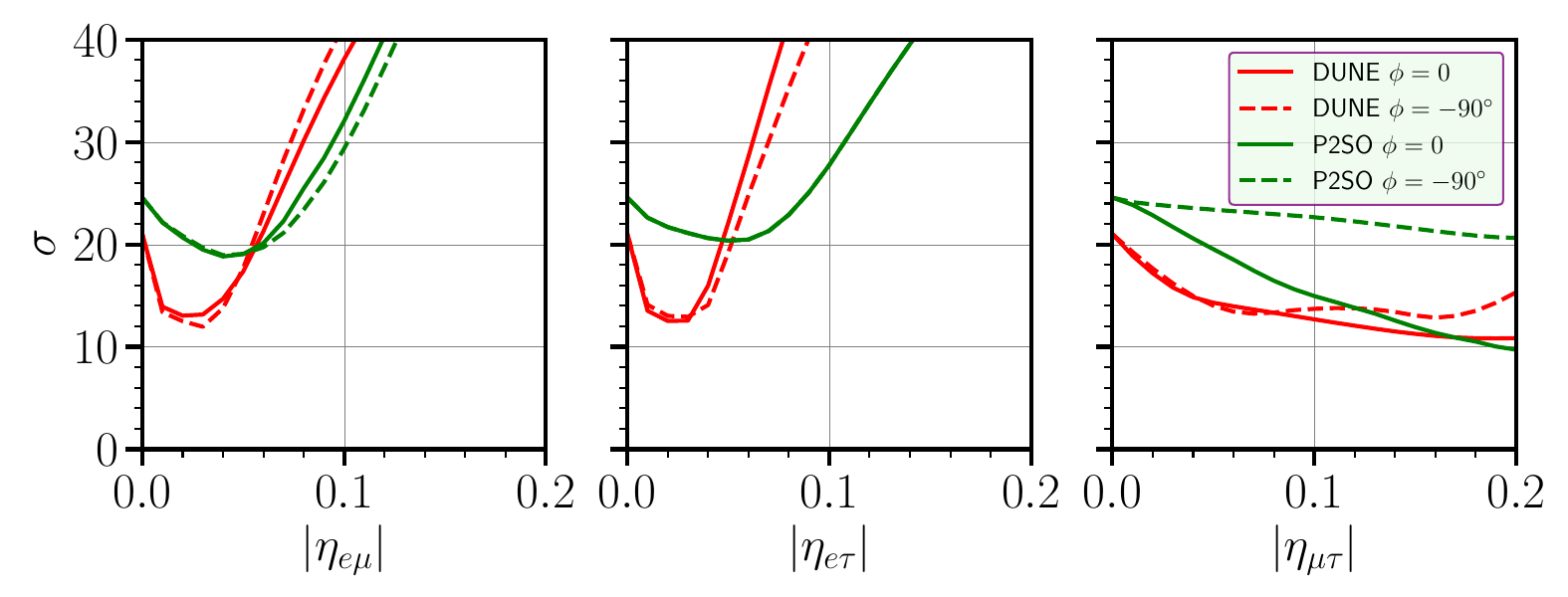}
    \caption{Variation of mass ordering sensitivity as a function of $|\eta_{\alpha\beta}|$ for DUNE and P2SO experiments for different value of true phases.}

    \label{fig:eta-MH}
\end{figure}
%%%%%%%%%%%%%
Figure~\ref{fig:eta-MH} displays the variation of mass ordering sensitivity as a function of true $|\eta_{\alpha\beta}|$.  The experiment's ability to reject the wrong mass ordering when determining the true mass hierarchy as normal ordering is shown in Fig.~\ref{fig:eta-MH}. We have done the analysis by considering two cases for the true value of $\phi_{\alpha \beta}$ as $0$ and $-90^\circ$. The left/middle/right panel shows the mass ordering sensitivities for the SNSI parameters $|\emu|/|\etau|/|\mutau|$, respectively. The red (green) curve represents the sensitivity for the DUNE (P2SO) experiment. Solid (dashed) curves are for $\phi^{\rm true}_{\alpha \beta} = 0 (-90^\circ)$.  The behavior of the curves with respect to $\eta_{\alpha \beta}$ is almost similar for the cases  $\emu$ and $\etau$. As $\eta_{\alpha \beta}$ increases, the sensitivity initially decreases and then it increases. For $\eta_{\mu\tau}$, the sensitivity decreases continuously. The behavior is similar for both DUNE and P2SO.

\begin{figure}[htbp!]
    \centering
 \includegraphics[width=55mm, height=52mm]{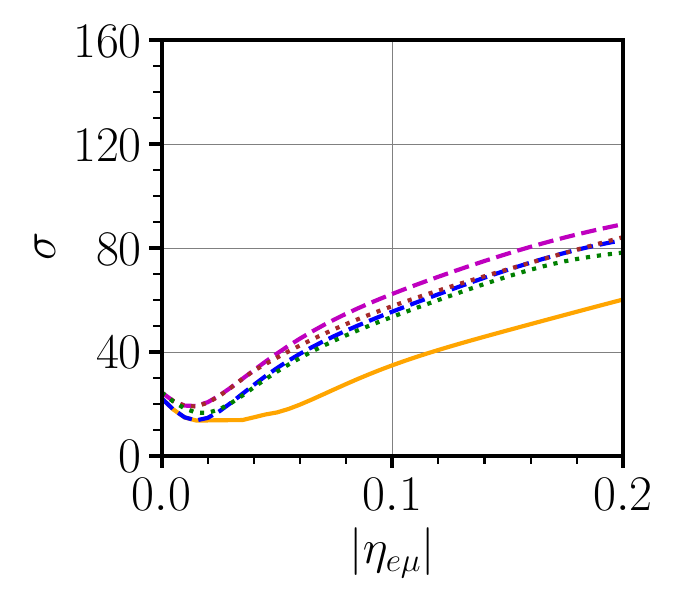}
    \includegraphics[width=52mm, height=52mm]{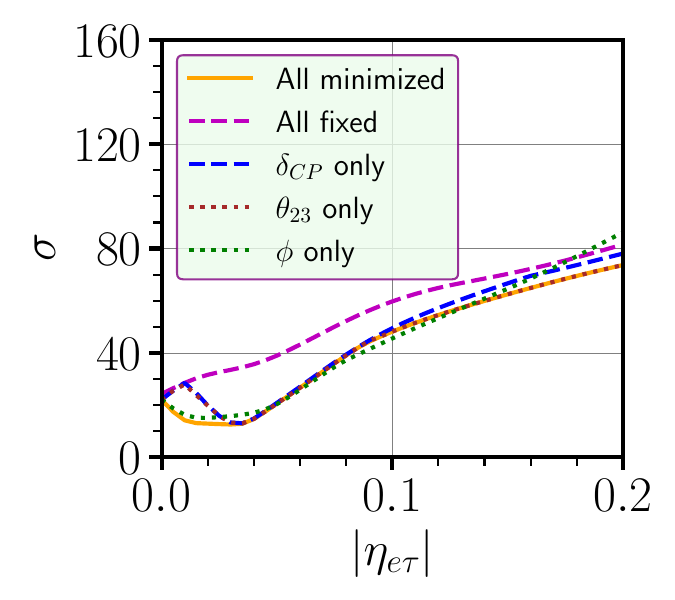}
    \includegraphics[width=52mm, height=52mm]{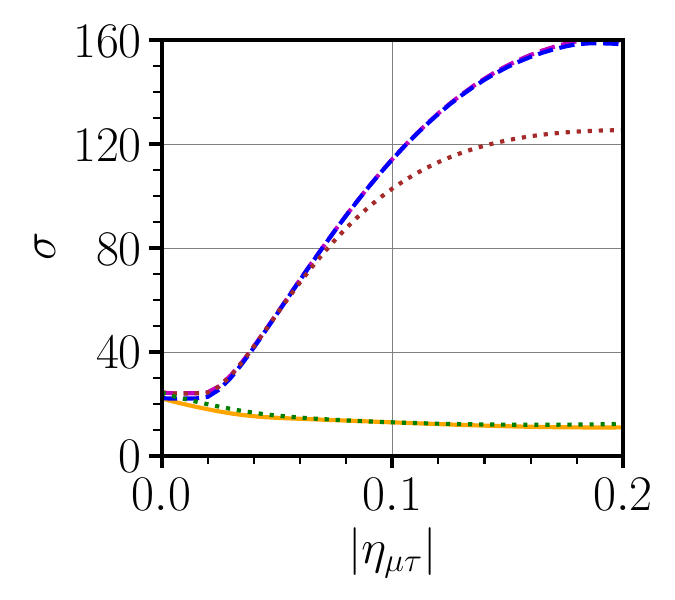}
        \caption{Variation of mass ordering sensitivity as a function of $|\eta_{\alpha\beta}|$ with different minimization conditions over parameters for $\phi^{\rm true}_{\alpha \beta}=0$ considering DUNE experiment. }
    \label{fig:eta-MH-2}
\end{figure}

To understand the behavior of the mass ordering curves in Fig.~\ref{fig:eta-MH-2}, we tried to look at the effect of minimization of the different oscillation parameters. In this figure, the left, middle and right panels are the sensitivities in presence of $\emu$, $\etau$ and $\mutau$, respectively. In each cases the true value of $\phi_{\alpha \beta}$ is considered to be $0$. In each panel, the orange curve is the scenario when minimization is done over all the oscillation parameters, while the magenta curve is the scenario without any minimization over any parameters. The blue/brown/green curve is obtained with minimizing over only $\delta_{\rm CP}$/$\theta_{23}$/$\phi_{\alpha\beta}$, respectively. This figure provides some important insights. For $\eta_{\mu\tau}$, if $\phi_{\mu\tau}$ is not minimized in the test, then the behavior of the curve will be the same as $\emu$ and $\etau$ i.e., sensitivity will first decrease as $\eta_{\mu\tau}$ increases and then it will increase. In this case,   because of the degeneracy with $\phi_{\alpha \beta}$, the behavior of the curve gets changed. For $\eta_{e\mu}$, we see that even when all the parameters are fixed, the behaviour of the curve remains the same, reflecting the fact that this behaviour does not arise due to the parameter degeneracy. For $\eta_{e\tau}$, one can see the effect of degeneracies associated with $\phi_{e\tau}$ and $\delta_{CP}$. The behavior of the mass ordering curves for  $\phi^{\rm true}_{\alpha \beta} = -90^\circ$ can be explained in a similar way.

\subsection{CP violation sensitivity}

%%%%%%%%%%%%%%%%%%%%%%%%%%%%%%%%%%%%%%%%%%%
\begin{figure}
    \centering

    \includegraphics[width=54mm, height=48mm]{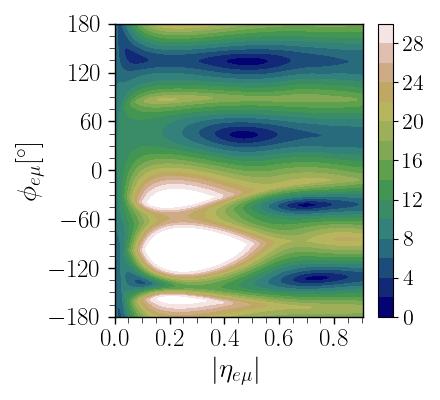}
    \includegraphics[width=54mm, height=48mm]{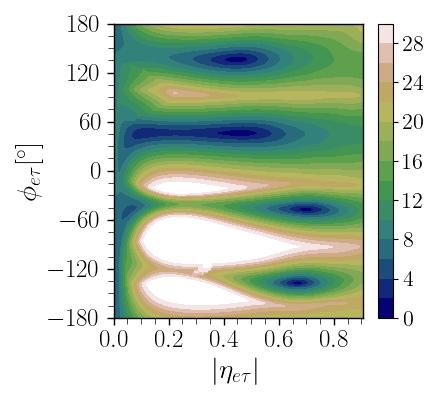}
    \includegraphics[width=54mm, height=48mm]{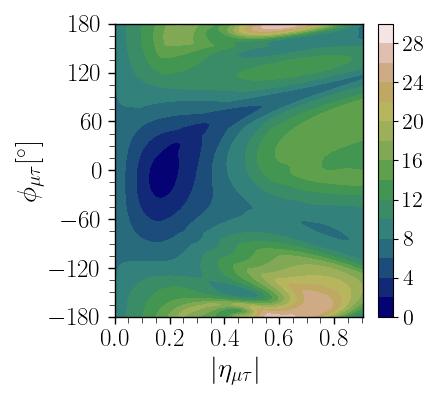}
    \includegraphics[width=54mm, height=48mm]{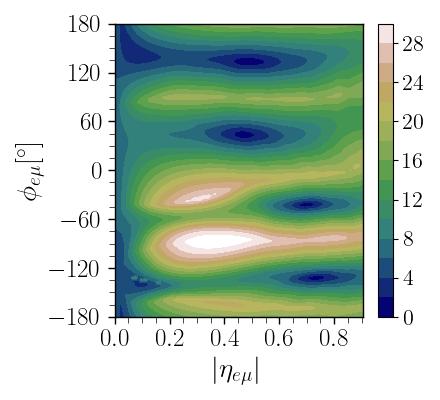}
    \includegraphics[width=54mm, height=48mm]{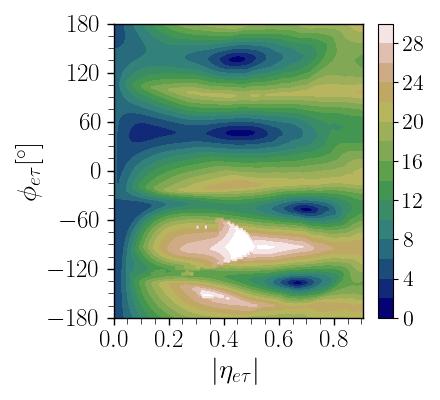}
    \includegraphics[width=54mm, height=48mm]{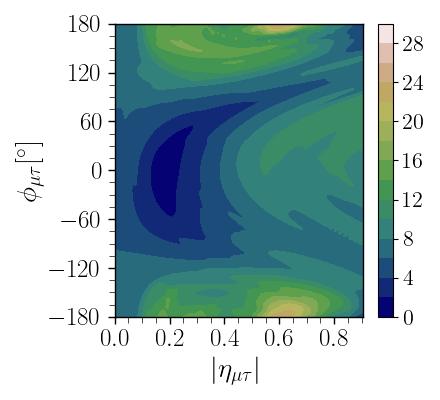}
    \caption{CP violation sensitivity as a function of $|\eta_{\alpha\beta}|$ and $\phi_{\alpha\beta}$ for DUNE (upper) and P2SO (lower) experiments.}

    \label{fig:CPVall}
\end{figure}

%%%%%%%%%%%%%%%%%%%%%%%%%%%%%%%%%%%%%%%%%%%
In Fig.~\ref{fig:CPVall}, we have shown CP violation (CPV) sensitivity for all possible true values of SNSI parameters $\eta_{\alpha\beta}$ and $\phi_{\alpha\beta}$. The upper (lower) panel is for DUNE (P2SO) experiment. CP violation sensitivity signifies the ability to exclude the CP conserving values of $\delta_{\rm CP}$. Here, we have considered the maximum CP violation scenario for $\delta_{CP}^{\rm true}$= $-90^\circ$. The color shades in each panel show the CPV sensitivity in terms of $\sqrt{\chi^2}$. It is very interesting to see that in the presence of all three off-diagonal SNSI parameters, the CP sensitivity of both the experiments can become extremely small. This is reflected by the combinations of some $|\eta_{\alpha\beta}| - \phi_{\alpha\beta}$ for which we obtain the dark blue regions. Table~\ref{tab:3 } shows the values of $|\eta_{\alpha\beta}|$ and $\phi_{\alpha\beta}$ for which the CP sensitivities of both experiments are almost lost.

\begin{table}[htbp!]
\begin{tabular}{|c|c|c|c|c|}
\hline
SNSI parameter & \multicolumn{2}{c|}{DUNE} & \multicolumn{2}{c|}{P2SO} \\
\cline{2-5}
 & Value & Phase ($\phi$) & Value & Phase ($\phi$) \\
\hline
$\eta_{e\mu}$  & 0.47 & $44^{\circ}$ & 0.47 & $44^{\circ}$ \\
  & 0.50 & $50^{\circ}$ & 0.49 & $49^{\circ}$ \\
  & 0.52 & $55^{\circ}$ & 0.51 & $54^{\circ}$ \\
  & 0.48 & $48^{\circ}$ & 0.47 & $47^{\circ}$ \\

\hline
$\eta_{e\tau}$ & 0.48 & $46^{\circ}$ & 0.45 & $46^{\circ}$ \\
  & 0.50 & $48^{\circ}$ & 0.47 & $47^{\circ}$ \\
  & 0.53 & $50^{\circ}$ & 0.50 & $49^{\circ}$ \\
  & 0.49 & $47^{\circ}$ & 0.46 & $46^{\circ}$ \\

\hline
$\eta_{\mu\tau}$ & 0.18 & $6^{\circ}$ & 0.18 & $8^{\circ}$ \\
\hline
\end{tabular}
  \caption{Values of SNSI parameters at which CP sensitivity is lost for DUNE and P2SO experiment.}
  \label{tab:3 }
\end{table}

\subsection{Ocatnt sensitivity}

\begin{figure}[htbp!]
    \centering
\includegraphics[width=168mm, height=62mm]{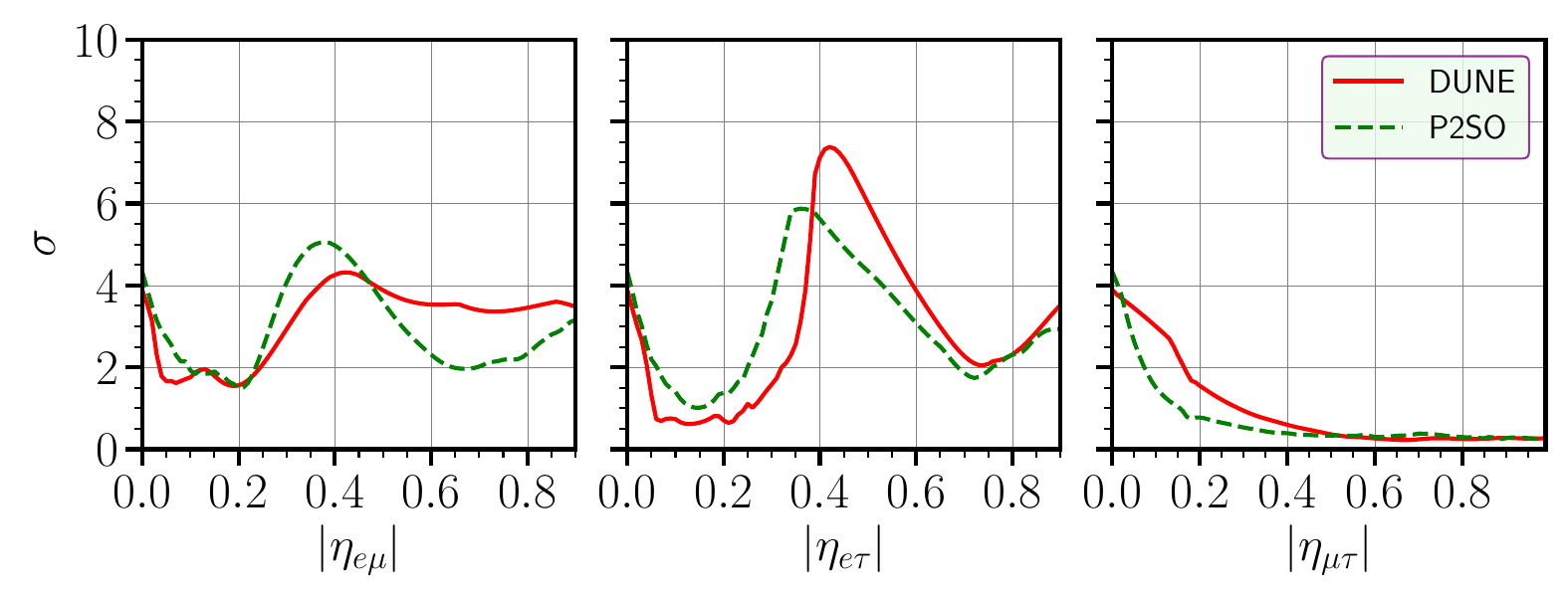}
        \caption{Octant sensitivity as a function of $|\eta_{\alpha\beta}|$ for DUNE and P2SO experiments. Here, the true octant considered to be in the LO.}
    \label{fig:octant}
\end{figure}

Figure~\ref{fig:octant} shows the octant sensitivity as a function of $|\eta_{\alpha\beta}|$ for the DUNE and P2SO experiments. Octant sensitivity signifies the  capability of the experiments to exclude the possibility of the wrong octant of $\theta_{23}$. We have obtained the sensitivity by taking into account the true octant of $\theta_{23}$ in the lower octant (LO) and higher octant (HO) in the test hypothesis. In the left/middle/right panel, we have shown the octant sensitivity in presence of $|\emu|/|\etau|/|\mutau|$ and red (green) curves are for DUNE (P2SO) experiment. For the parameters $|\eta_{e\mu}|$ and $|\eta_{e\tau}|$, we observe a peak around 0.4 whereas for $|\eta_{\mu\tau}|$, the sensitivity decreases steadily as $\eta$ increases. These behaviours are mainly governed by the parameter degeneracies associated with the parameter $\phi_{\alpha\beta}$.

\subsection{Precision of $\sin^2\theta_{23}$ and $\Delta m_{31}^2$}

\begin{figure}[htbp!]
    \centering

    \includegraphics[width=54mm, height=54mm]{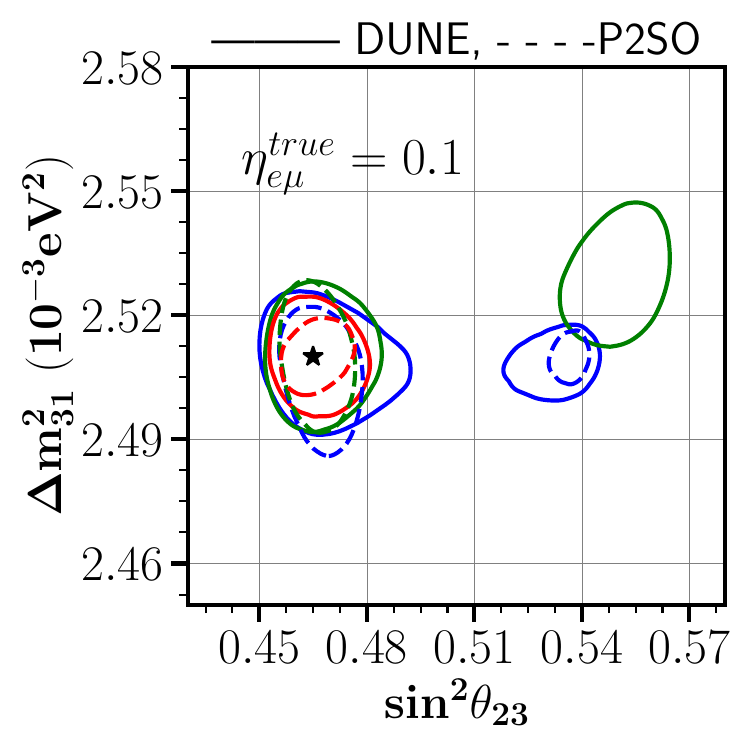}
    \includegraphics[width=54mm, height=54mm]{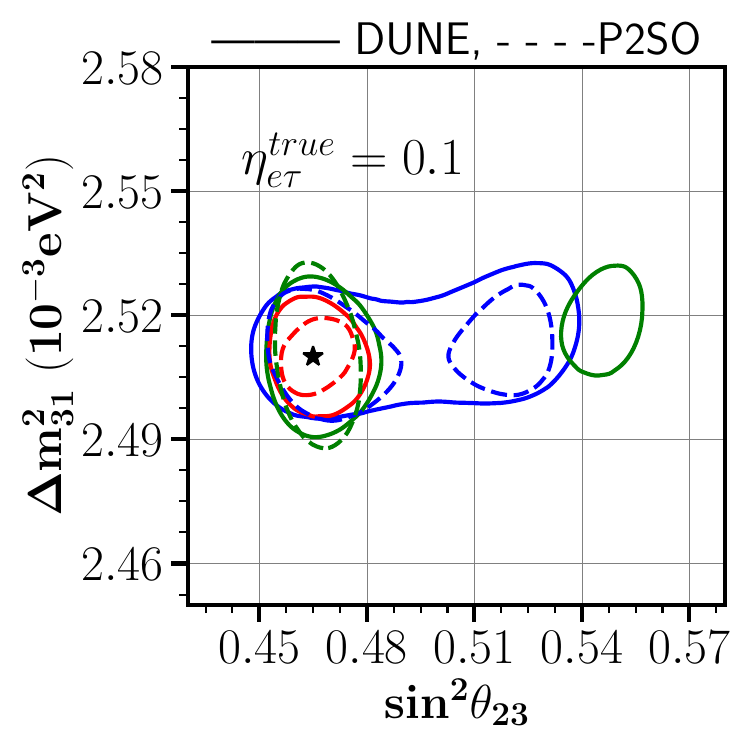}
    \includegraphics[width=54mm, height=54mm]{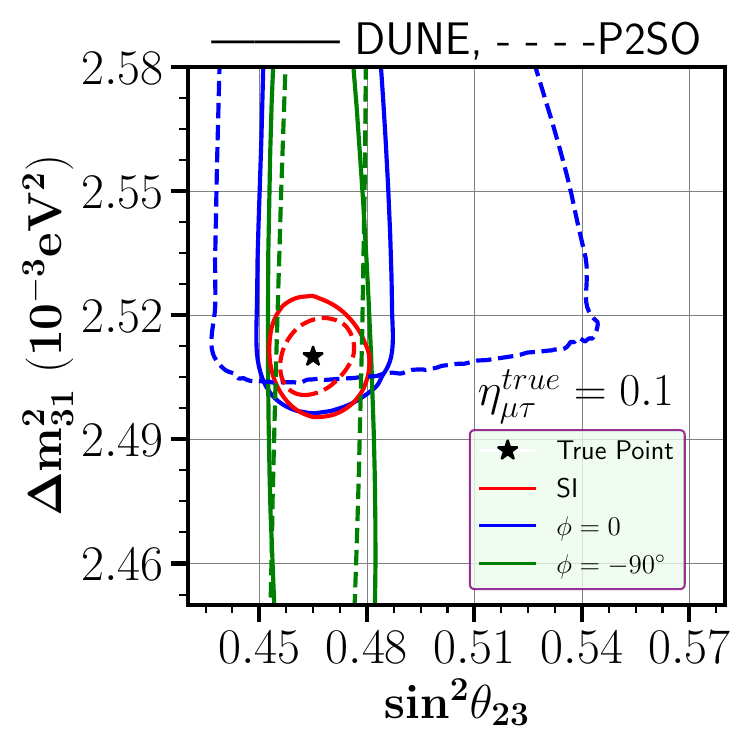} \\
    \includegraphics[width=54mm, height=54mm]{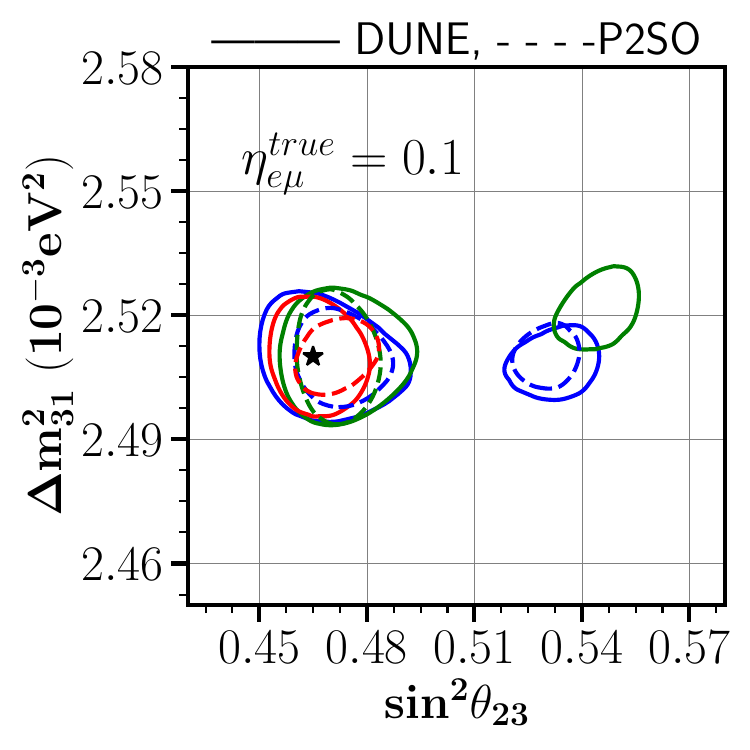}
    \includegraphics[width=54mm, height=54mm]{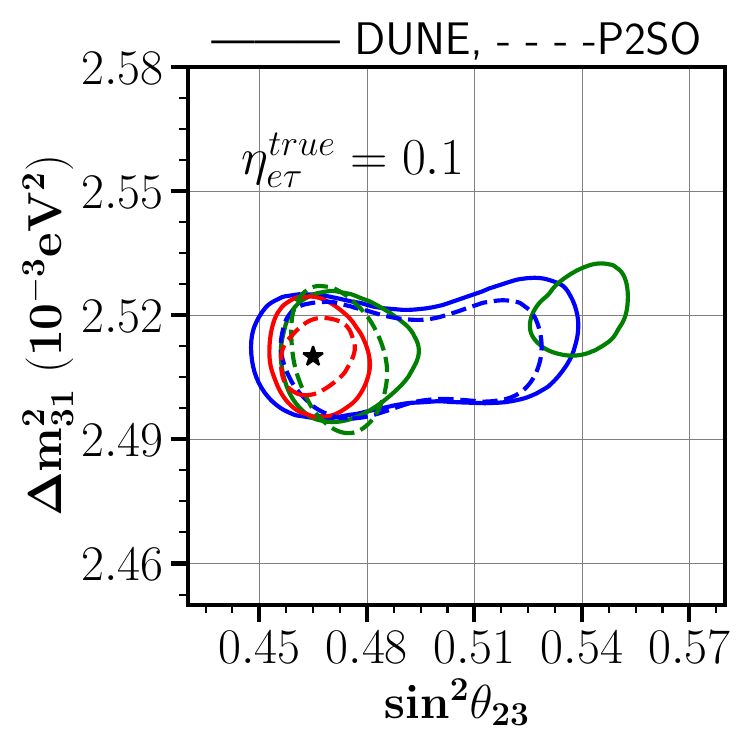}
    \includegraphics[width=54mm, height=54mm]{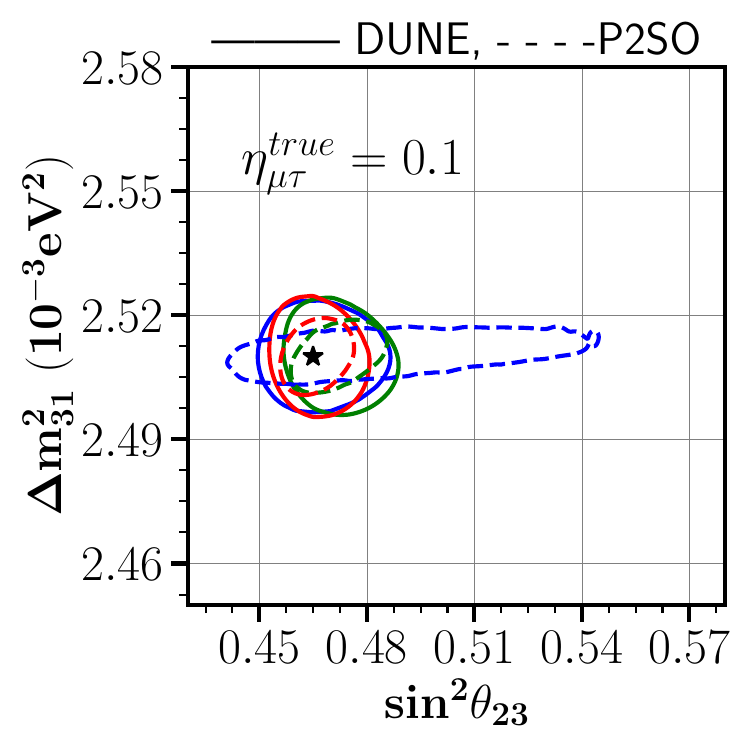}
        \caption{Allowed parameter space between $\sin^2\theta_{23}$-$\Delta m_{31}^2$ plane for DUNE and P2SO experiments at $90 \%$ C.L. Top row: with marginalization over $\phi_{\alpha \beta}$, bottom row: without marginalization over $\phi_{\alpha \beta}$. }
    \label{fig:eta-th23-ldm}
\end{figure}
%%%%%%%%%%%%%%%%%%%%%%%%%%%%%%%%%%%%%%%%%%%%

In this section, we aim to show the effect of SNSI parameters on the precision measurement of $\sin^2\theta_{23}$ and $\Delta m_{31}^2$. Figure \ref{fig:eta-th23-ldm} shows the allowed parameter space between $\sin^2\theta_{23}$ and $\Delta m^2_{31}$ in the presence of SNSI parameters as well as in the standard interaction (SI) case at 90$\%$ C.L.. We have considered a benchmark value $|\eta_{\alpha\beta}|$=0.1 in both the true and the test scenario for all the cases of $|\eta_{\alpha\beta}|$. Furthermore, we looked into two scenarios for the true value of $\phi_{\alpha\beta}$, which are $-90^\circ$ and $0$. In each row, left/middle/right panel is for $|\eta_{e\mu}|/|\eta_{e\tau}|/|\eta_{\mu\tau}|$, respectively. The values of $\Delta m^2_{31}$ in y-axis correspond to the current $3 \sigma$ range of this parameters. In each panel, solid and dashed  contours are for DUNE and P2SO, respectively. The red contour represents the allowed region in SI case, while the green/blue contour symbolizes the allowed region for the true value of $\phi_{\alpha\beta}$ as $-90^\circ$/$0$. The true point is represented by the black star in each plot. 

First let us focus on the upper row where the test value of $\phi_{\alpha\beta}$ is varied over all the allowed values, considering that exact value $\phi_{\alpha\beta}$ is not known. In the presence of SNSI parameters, the allowed parameter space increases in size compared to the SI case, which implies that the precision of these parameters gets worsen in the presence of SNSI. The deterioration of sensitivity along the $\Delta m^2_{31}$ direction is not very significant for either DUNE or P2SO in the presence of $\eta_{e\mu}$ and $\eta_{e\tau}$. However, both experiments show a loss of sensitivities in the precision of $\theta_{23}$ which is evident from the existence of allowed regions in the wrong octant. For $\eta_{\mu\tau}$, the precision of both $\Delta m^2_{31}$ and $\theta_{23}$ gets deteriorated for P2SO whereas DUNE sees only a deterioration in $\Delta m^2_{31}$.  The deterioration of the $\theta_{23}$ precision is mainly of the worsening of the octant sensitivity at these values of $\eta$ which we can see from Fig.~\ref{fig:octant}.

To understand the deterioration of the precision of $\Delta m^2_{31}$ in presence of $\eta_{\mu\tau}$, let us focus on the lower row of Fig.~\ref{fig:eta-th23-ldm} where we have kept $\phi_{\alpha\beta}$ to be fixed to its true values. Here interestingly we see that the the precision of $\Delta m^2_{31}$ became close to the standard value for $\eta_{\mu\tau}$ implying the fact that the the phase $\phi_{\mu\tau}$ was solely responsible for the deterioration of the precision of $\Delta m^2_{31}$.

 Here we would like to mention that the inclusion of atmospheric neutrino oscillation data with long-baseline experiments can further improve the precision of these parameters.

%%%%%%%%%%%%%%%%%%%%%%%%%%%%%%%%%%%%%%%%%

\section{Summary and conclusion}
\label{conclusion}

In this paper, we have studied the impact of the off-diagonal SNSI parameters in the future long-baseline neutrino oscillation experiments DUNE and P2SO.  In our analysis, we have found  that $\Delta m^2_{31}$ does not play a significant role in deriving bounds on the SNSI
off-diagonal parameters, in contrast to the diagonal parameters, as discussed in Ref.~\cite{Singha:2023set}. For the parameters $\eta_{e\mu}$ and $\eta_{e\tau}$, SNSI can be fitted with the standard three flavour scenario,  with a value of  $\Delta m^2_{31}$ within its current $3 \sigma$ range.  However, for $\eta_{\mu\tau}$, values of $\Delta m^2_{31}$ beyond its current $3 \sigma$ values can fit SNSI with the standard scenario. For all these three off-diagonal parameters, we have identified a degeneracy where the standard three flavour scenario mimics the SNSI scenario even for non-zero values of $\eta_{e\mu}$, $\eta_{e\tau}$ and $\eta_{\mu\tau}$. We have also found out that the bounds on the SNSI parameters also depend on the phases associated with them. Among the three parameters, the bound on $\eta_{e\mu}$ is strongest and the bound on $\eta_{\mu\tau}$ is weakest. The sensitivities of both experiments are very similar in this regard. 

Next, we tried to understand how the sensitivities of these experiments get affected if one assumes that the off-diagonal SNSI exists in Nature. Regarding the mass ordering sensitivity, we have shown that as $\eta_{\alpha \beta}$ increases from zero, the sensitivity initially decreases and then increases for the parameters $\eta_{e\mu}$ and $\eta_{e\tau}$, whereas for the parameter $\eta_{\mu\tau}$, it decreases continuously. We have explained this behaviour from the argument of degeneracies among different parameters. For CP sensitivity, the conclusions for the off-diagonal parameters are very similar as the diagonal parameters i.e., for some combinations of $|\eta_{\alpha \beta}|$ and $\phi_{\alpha \beta}$, the CP sensitivity of both the experiments can be very small or even they can vanish. Regarding octant sensitivity, for the parameters $|\eta_{e\mu}|$ and $|\eta_{e\tau}|$, we observe a peak around 0.4 whereas for $|\eta_{\mu\tau}|$, the sensitivity decreases steadily as $\eta$ increases. Finally, with respect to the precision of the atmospheric mixing parameters $\Delta m^2_{31}$ and $\theta_{23}$, we find that the presence of off-diagonal SNSI leads to a reduction in their sensitivity compared to the standard scenario, specially for $\theta_{23}$. When the phase of the complex $\eta$ parameter is fixed in the test spectrum, the deterioration in the precision of $\Delta m^2_{31}$ for the case $\eta_{\mu\tau}$ gets resolved.

In conclusion, we would like to emphasize that if a non-zero off-diagonal SNSI exists in nature, it will alter the phenomenology of the long-baseline experiments in a very non-trivial way. For example, depending on the values of these parameters, they can either completely mimic the standard scenario or  can wash out their CP sensitivity. Therefore, it is very important to look for the existence of this theory in the current and future data of the neutrino oscillation experiments. 

\section*{Acknowledgments}

The work of MG has been in part funded by (i) Ministry of Science and Education of Republic of Croatia grant No. KK.01.1.1.01.0001, (ii) SNSF, HRZZ and NRDIO under grant MAPS IZ11Z0$\_$230193 and (iii) European Union under the NextGenerationEU Programme. Views and opinions expressed are however those of the author(s) only and do not necessarily reflect those of the European Union. Neither the European Union nor the granting authority can be held responsible for them. SKP would like to acknowledge University Grants Commission for the  NFOBC fellowship. DKS would like to acknowledge Prime Minister's Research Fellowship, Govt. of India. RM (Rudra Majhi) would like to acknowledge Odisha State Higher Education Council, Govt. of Odisha for the support under Mukhyamantri Research and Innovation (MRIP)-2024 (24EM/PH/102).  RM  acknowledges the support from University of Hyderabad  through IoE project grant no. RC1-20-012.
 
\bibliography{SNSI-ref}
\end{document}